\documentclass[onecolumn,nofootinbib,amsmath,amssymb,amsfonts,aps,prfluids]{revtex4-2}

\usepackage{graphicx}
\usepackage{dcolumn}
\usepackage[dvipsnames]{xcolor}

\usepackage{hyperref}
\hypersetup{colorlinks=true,linkcolor=blue,urlcolor=blue,citecolor=blue}

\usepackage[executivepaper,margin=0.75in]{geometry}

\begin{document}

\title{The influence of incompressible surfactant on drag in flow along an array of gas-filled grooves}

\author{Tobias Baier}
\email[]{baier@nmf.tu-darmstadt.de}
\affiliation{Fachbereich Maschinenbau, Technische Universit\"at Darmstadt, 64287 Darmstadt, Germany}



\begin{abstract}
Surfactants can have a detrimental effect on the drag reduction in shear flow over superhydrophobic surfaces in Cassie state. While surfactant-free gas-liquid interfaces are often well approximated as shear-free, surfactants can impede the flow by stacking up in front of obstacles. We study shear-flow along an array of narrow gas-filled grooves of finite length embedded in an otherwise planar surface, with the gas-liquid interface protruding slightly above or below the plane. Assuming immiscible surfactants forming an incompressible, inviscid surfactant phase at the gas-liquid interfaces we employ a recently proposed model [Baier and Hardt, J.Fluid Mech., 949 (2022)] for addressing this situation. Using a domain perturbation technique together with the Lorentz reciprocal theorem we obtain the slip length characterizing the flow over such surfaces to second order in the maximal interface deflection as a small parameter. We find that within the range of moderate interface deflections studied, the slip length for flow over such surfaces is negative (positive) for surfaces protruding above (below) the surface and is much smaller than for flow over a corresponding surfactant-free interface. Thus, contrary to expectations of reduced drag in flow over superhydrophobic surfaces in Cassie state, surfactant covered interfaces can even be detrimental for drag reduction in the limit where surfactants act as an effectively incompressible surface-fluid. This has important implications for the appropriate design of superhydrophobic surfaces for reducing flow resistance.
\end{abstract}

\maketitle

\section{Introduction}\label{sec:intro}

Superhydrophobic surfaces containing gas-filled cavities are a prototypical example for surfaces promising the reduction of drag in near-wall flow of liquids \citep{Rothstein_2010, Lee_2016}. This promise is supported when the gas-liquid interface can be considered nearly stress-free due to the low gas viscosity compared to the viscosity of the liquid. An illustrative example is an array of gas filled grooves embedded in a planar surface as sketched in figure \ref{fig:sketch}(a) over which a liquid is forced to flow by the motion of a plate moving at some distance in parallel to the structured surface. For a flat gas-liquid interface an analytical description is known \citep{Philip_1972, Lauga_2003}, which was later extended to account for deformations of the interface \citep{Sbragaglia_2007, Teo_2010, Crowdy_2010, Crowdy_2015, Crowdy_2016, Schnitzer_2017, Kirk_2018} or for describing liquid infused surfaces, where the gas is replaced by a liquid immisscible with the main liquid \citep{Schonecker_2014, Asmolov_2018}. Experimentally, the drag reduction is readily characterised using a viscosimeter by measuring the stress in a shear flow over such surfaces in setups similar to the one sketched in figure \ref{fig:sketch}a and it is customary to introduce an effective slip length as a more tangible measure for drag reduction than stress itself \citep{Rothstein_2010}.

It has long been known that even small amounts of surfactants can influence the flow at gas-liquid interfaces by stacking up in front of obstacles \citep{Merson_1965, Harper_1992} or aggregating at the downstream hemisphere of bubbles rising in a liquid \citep{Sadhal_1983}, to give just two examples. For this reason they have been aptly described as 'hidden variables' influencing fluid flow \citep{Manikantan_2020}. Correspondingly, their impact was also observed in flow over superhydrophobic surfaces \citep{Kim_2012, Bolognesi_2014, Schaffel_2016, Peaudecerf_2017, Song_2018, Li_2020}, where surfactants stack up at downstream edges of gas-filled cavities or upstream of pillars piercing the interface. Attempts to describe such flows have focused mostly on rectangular gas-filled cavities parallel or tangential to the flow direction \citep{Gaddam_2018, Landel_2020, Baier_2021, Temprano_2023} or their counterparts for liquid infused surfaces \citep{Sundin_2022}. Sufficient surfactant coverage can drastically reduce the drag-reduction properties of superhydrophobic surfaces when Marangoni stresses within the interface due to gradients in surface tension become of the same order of magnitude as viscous stress applied to the surface. For large Marangoni number, Langmuir monolayers of insoluble surfactant molecules can even become effectively incompressible \citep{Manikantan_2020}, when the surface pressure within in the surface film effectively inhibits compression. In this case the gas-liquid interface can become partially or completely immobilized \citep{Peaudecerf_2017, Baier_2021, Mayer_2022} in case of flat interfaces. However, generally the pressure in the liquid and the gas trapped in the cavities are not necessarily the same, implying a curved interphase between them. In such situations recirculation zones have been observed experimentally at the curved gas-liquid interface when covered by a nearly incompressible surfactant film \citep{Song_2018, Li_2020}. Recently, we have proposed a model describing the flow over a long, narrow cavity covered by an incompressible surfactant phase \citep{Baier_2022}, which compares favourably with the flow pattern observed in the experiments by \citet{Song_2018}. The non-vanishing interface velocity observed in this case naturally raises the question whether such surfaces are still suitable for drag reduction despite the presence of surfactants. Here we therefore extend the theoretical analysis of flow over a single gas-filled groove covered by an incompressible surfactant phase to flow over an array of such grooves in order to investigate the effective slip expected in this situation. 
Note that in the limit of large groove separation this was recently also investigated by \citet{Rodriguez_2023} using a superposition of the velocity fields for flow over single grooves to obtain the shear stress far from an interface with a dilute array of grooves covered by incompressible surfactant. In the present analysis no restriction is made for the separation between grooves, allowing interaction between the velocity fields at neighboring grooves.

We will proceed as in \citet{Baier_2022} using a domain perturbation technique to obtain the velocity field as an expansion in the interface-deflection as a small parameter. However, here we shall mainly be interested in integral properties such as the stress on a moving surface driving the forcing shear flow or the effective slip length for shear flow over a surface containing such grooves. As we will see, the Lorentz reciprocal theorem allows us to obtain these quantities to second order in the interface deflection when the velocity field is obtained to first order only. The analytical results are complemented by numerical calculations.

\section{Modelling an incompressible surface fluid}

\begin{figure}
	\centering
	\includegraphics[width=0.9\textwidth]{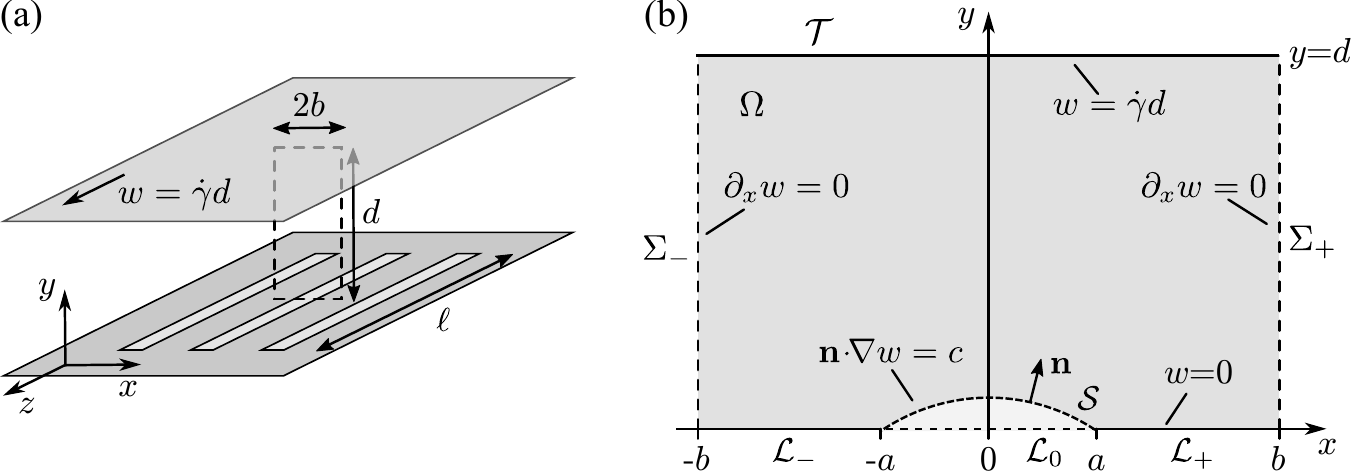}
	\caption{\label{fig:sketch} (a) Sketch of the configuration under investigation with a liquid in the gap between two parallel plates separated by a distance $d$. The lower surface contains an evenly spaced array of parallel long, narrow gas-filled cavities of width $2a$ and length $\ell$ (with $\ell \gg d, a$) at a pitch of $2b$. The upper plate moves at velocity $w=\dot{\gamma}d$ in the $z$-direction, driving a Couette flow along the grooves. 
	The planar region of interest, indicated by the dashed rectangle, is sufficiently far away from the ends of the grooves, such that the deflection $h(x)$ of the gas-liquid interface can be considered independent of $z$. 
	(b) Region of interest in the $x$--$y$-plane, straddling a single groove. The gas-liquid interface, ${\cal S}$, is assumed to have the shape of a circular arc and is laden with an insoluble, incompressible surfactant. The lower walls ${\cal L}_\pm$ are no-slip surfaces while the upper wall ${\cal T}$ moves at velocity $w=\dot{\gamma}d$ in the direction normal to the plane of the figure. By symmetry, the sides $\Sigma_\pm$ are no-shear surfaces.
	}
\end{figure}

A sketch of the investigated configuration is shown in figure \ref{fig:sketch}(a). An incompressible Newtonian liquid of viscosity $\mu$ fills the gap between two parallel plates separated by a distance $d$. The lower plate lies in the $x$--$z$ plane and contains a periodic array of long, narrow gas-filled grooves of width $2a$ and length $\ell$ at a center-to-center separation of $2b$. The lower plate is at rest, while the unstructured upper plate moves at a constant velocity $w=\dot{\gamma}d$ in $z$-direction parallel to the grooves, driving a Couette flow between the plates. We assume a small difference in pressure between the gas in the groove and in the fluid, such that the gas-liquid interface is deflected slightly above or below the $x$--$z$ plane. For $\ell \gg d$ and considering a region far from the ends of the grooves, the deflection, $y=h(x)$, of the gas-liquid interface has the form of a circular arc and can be considered independent of the position $z$ along the grooves such that the flow velocity $\boldsymbol{w}=w(x,y) \boldsymbol{e}_z$ becomes uni-directional and translationally invariant along the grooves. We can then pick a region of interest $\Omega$ as one unit cell with width $2b$ perpendicular to and straddling one of the grooves, indicated by the dashed line in figure \ref{fig:sketch}(a) and shown in the $x$--$y$ plane in figure \ref{fig:sketch}(b). Under these conditions the Navier-Stokes equation governing the velocity field reduces to the Laplace equation
\begin{equation}\label{eq:dimLaplace}
  \boldsymbol{\nabla}^2 w(x,y) = 0,\quad \text{in }\Omega
\end{equation}
with periodic boundary conditions
\begin{equation}\label{eq:dimPeriodocity}
  \partial_x w(\pm b, y) = 0,\quad \text{on }\Sigma_\pm
\end{equation}
on the sidewalls and no-slip Dirichlet conditions
\begin{align}
  w(x,0) &= 0,\quad\;\, \text{on }{\cal L}_\pm \label{eq:dimDirichletLowerWall} \\
  w(x,d) &= \dot{\gamma} d,\quad \text{on }{\cal T} \label{eq:dimDirichletUpperWall}
\end{align}
on the solid sections of the stationary lower wall and the translating upper wall. The gas-liquid interface ${\cal S}$ is assumed to have the form of a circular arc and covered by an incompressible inviscid surface fluid. Since the grooves are of finite extend, there is no net flow of the surface fluid along the groove such that conservation of its mass demands
\begin{equation}\label{eq:dimIntegralMassConservationSurfaceFluid}
  \int_{\cal S} w(x,y) ds = 0,
\end{equation}
where the integral is along the arc of the interface lying in the domain of interest. Thus, when the fluid follows the main flow direction on some parts of the gas-liquid interface, the flow direction must be in the opposite direction on other parts of the interface. The driving force of this flow reversal is the surface pressure (or Marangoni stress) in the surface fluid opposing compression. In \cite{Baier_2022} it was shown that the tangential stress balance on the interface becomes
\begin{equation}\label{eq:dimStressInterface}
\mathbf{n}\cdot (\mu \boldsymbol{\nabla} w) = \partial_z \Pi = c, \quad \text{on } {\cal S}.
\end{equation}
with a constant gradient in surface pressure $\partial_z \Pi(z) = c$ opposing compression of the surface fluid by the viscous stress from the liquid, while neglecting the influence of the gas in the groove due to its low viscosity compared to the liquid. The so far undetermined stress $c$ in \eqref{eq:dimStressInterface} is fixed by the integral mass conservation condition \eqref{eq:dimIntegralMassConservationSurfaceFluid} on the interface and similar to the pressure acting as a Lagrange multiplier for the incompressibility condition of stationary Stokes flow derived from minimising energy dissipation \citep{Ozanski_2017}, it can be viewed as a Lagrange multiplier ensuring incompressibility of the surface fluid \citep{Baier_2022}. For further justification of this model and a discussion of it limits of applicability we refer the reader to the appendix of \citep{Baier_2022}.
Briefly, the response of the interfacial surfactant concentration $\Gamma$ to an applied tangential shear stress $\tau$ along the interface can be characterized by the Gibbs elastic modulus $E = \Gamma (\partial \Pi/\partial \Gamma) = -\Gamma (\partial \gamma/\partial \Gamma)$, indicating the change in interfacial pressure $\Pi$ or surface tension $\gamma$ with interfacial surfactant concentration $\Gamma$ \citep{Manikantan_2020}. At equilibrium, an applied shear stress $\tau$ along a groove of length $\ell$ is compensated by a corresponding Marangoni stress $\Delta\Pi/\ell \simeq (E_0/\ell) (\Delta \Gamma/\Gamma)$ due to a variation $\Delta \Gamma$ in surfactant density over a length scale $\ell$. Thus, for large Marangoni number, $\text{Ma}=E/(\tau \ell)$, the interfacial surfactant phase becomes virtually incompressible, $\Delta \Gamma/\Gamma \ll 1$. Additionally, since concentration gradients are small, interfacial diffusion can be neglected and surfactant transport is dominated by convection (provided the corresponding P{\'e}clet numbers are not too small), such that equation \eqref{eq:dimIntegralMassConservationSurfaceFluid} is an integral statement of the equation of continuity for surfactant transport in the limit of large Marangoni numbers, while equation \eqref{eq:dimStressInterface} reflects the balance between Marangoni stress and viscous stress.

It is easy to find a solution to the above flow for vanishing deflection $h(x) = 0$ of the interface. We note that Couette flow between parallel plates, 
\begin{equation}\label{eq:dimCouetteFlat}
    w_0(x,y)=\dot{\gamma}y,
\end{equation}
has a constant shear rate $\dot{\gamma}$ throughout and therefore solves \eqref{eq:dimLaplace} with boundary conditions \eqref{eq:dimPeriodocity}-\eqref{eq:dimStressInterface} at an interfacial shear stress $c=\dot{\gamma}/\mu$. The interface thus becomes completely immobilized in this case. As edge cases in the limit of large surfactant coverage, this situation was already studied previously, \citep{Peaudecerf_2017, Mayer_2022,Baier_2021}, and we recognize the incompressiblility condition used here as the limiting case for flow at large Marangoni number \citep{Baier_2022}. 

\subsection{Dimensionless formulation and domain perturbation}

Using the scale $a$ for length and $u=a\dot{\gamma}$ for velocity, we introduce the dimensionless coordinates $(X,Y)=(x/a,y/a)$ and a dimensionless velocity $W(X,Y)=w(aX,aY)/(a\dot{\gamma})$ which obeys the Laplace equation \eqref{eq:dimLaplace}
\begin{equation}\label{eq:noDim_Laplace}
    \widetilde{\boldsymbol{\nabla}}\vphantom{\nabla}^2 W(X,Y) = 0,
\end{equation}
where $\widetilde{\boldsymbol{\nabla}}$ is the gradient in the dimensionless coordinates $(X,Y)$. With $B=b/a$ and $D=d/a$, the periodic boundary conditions \eqref{eq:dimPeriodocity} and Dirichlet conditions \eqref{eq:dimDirichletLowerWall}, \eqref{eq:dimDirichletUpperWall} on the solid wall sections become
\begin{alignat}{2}
W(X,0) &= 0, &\quad& 1 \leq |X| \leq B, \label{eq:noDim_DirichletLowerWall}\\
W(X,D) &= D, &&\ 0 \le |X| \le B,		\label{eq:noDim_DirichletUpperWall}\\
\partial_X W(\pm B, Y) &= 0, && 0 \le Y \le D. \label{eq:noDim_Periodicity} 
\end{alignat}
Parametrising the interface ${\cal S}$ by $Y=H(X)=h(aX)/a$, the integral mass conservation of the surface fluid \eqref{eq:dimIntegralMassConservationSurfaceFluid} and stress condition \eqref{eq:dimStressInterface} at the interface become
\begin{alignat}{2}
0 &= \int_{-1}^1 W(X,H(X))\sqrt{1+\left(H'(X)\right)^2} dX,&&\quad |X|<1,  \label{eq:noDim_IntegralMassConservationSurfaceFluid} \\
C &=\frac{(-\partial_X H(X)) \partial_X W(X,Y)+\partial_Y W(X,Y)}{\sqrt{1+(\partial_X H(X))^2}}\bigg|_{Y=H(X)},&&\quad |X|<1,  \label{eq:noDim_StressInterface}
\end{alignat}
where $C=c/(\mu\dot{\gamma})$ is the dimensionless stress along the interface. 

We characterize the interface by its maximal protrusion $\varepsilon = H(0) = h(0)/a$ above the $x$--axis, such that it is described as a segment of a circle of radius $r/a=|\varepsilon+\varepsilon^{-1}|/2$ and center at $Y = y/a = (\varepsilon-\varepsilon^{-1})/2$ on the $y$--axis. For small deflections, the interface is then described, up to second order in $\varepsilon$, by
\begin{equation} \label{eq:perturbationInterface}
    H(X) = \varepsilon H_1(X),\qquad H_1(X) = 1-X^2.
\end{equation}
In order to obtain the velocity field we proceed as in \citet{Baier_2022} by performing a domain perturbation \citep{Leal_2007} in the dimensionless deflection $\varepsilon$ with an expansion of the velocity field $W$ and tangential shear stress $C$ around the solution \eqref{eq:dimCouetteFlat} for a flat interface to second order in $\varepsilon$
\begin{align} \label{eq:perturbationExpansion}
W(X,Y) = Y + \varepsilon W_1(X,Y) + \varepsilon^2 W_2(X,Y), \qquad
C = 1 + \varepsilon C_1 + \varepsilon^2 C_2.
\end{align}
Inserting the expansions \eqref{eq:perturbationInterface} and \eqref{eq:perturbationExpansion} into the boundary conditions \eqref{eq:noDim_IntegralMassConservationSurfaceFluid} and \eqref{eq:noDim_StressInterface} on ${\cal S}$ leads to their projection onto the segment ${\cal L}_0$ of the real axis up to second order in $\varepsilon$ as
\begin{equation}\label{eq:epsIntBC_arc}
\begin{split}
0 = \int_{-1}^1 \left[ \varepsilon \left(W_1(X,0) + H_1(X) \right)
	+ \varepsilon^2 \left(W_2(X,0) + H_1(X)\partial_Y W_1(X,0) \right)
	\right] dX
\end{split}
\end{equation}
and
\begin{equation}\label{eq:epsBC_arc}
\begin{split}
\varepsilon C_1 + \varepsilon^2 C_2 = \varepsilon \partial_Y W_1(X,0) + \varepsilon^2 \left[ \partial_Y W_2(X,0) - \partial_X \big(H_1(X)\partial_X W_1(X,0)\big) - \tfrac{1}{2}\big(\partial_X H_1(X)\big)^2 \right].
\end{split}
\end{equation}
The domain perturbation thus considers a projection of the boundary conditions on ${\cal S}$ onto the segment ${\cal L}_0$ of the $x$-axis and the velocity field can be obtained order by order in $\varepsilon$ by solving the Laplace equation \eqref{eq:noDim_Laplace} for each $W_i(X,Y)$ in the rectangle $\Omega_0 = \{(X,Y)| -B \leq X \leq B, 0 \leq Y \leq D\}$ using the projected boundary conditions on ${\cal L}_0$.

We note that the integral boundary condition \eqref{eq:epsIntBC_arc} fixes the average velocity at order $\varepsilon^2$ on the real axis once the velocity at order $\varepsilon^1$ is known. As we will see in section \ref{sec:Stress_EffectiveSlip}, this is enough for determining the average shear rate on the moving wall to order $\varepsilon^2$, allowing us to restrict evaluating the velocity field to order $\varepsilon^1$.

\subsection{Velocity field}\label{seq:velocity}

The Laplace equation \eqref{eq:noDim_Laplace}, the condition of periodicity \eqref{eq:noDim_Periodicity} and the Dirichlet boundary condition \eqref{eq:noDim_DirichletLowerWall} at the lower wall apply to all $W_i(X,Y)$. Since $W(X,D)$ and $W_0(W,D)$ fulfill the Dirichlet boundary condition \eqref{eq:noDim_DirichletUpperWall} at the upper wall, the corresponding condition for $W_1$ becomes
\begin{equation}
    W_1(X,D) = 0,\quad 0\le |X| \le B.
\end{equation}
These conditions are accompanied by the stress condition \eqref{eq:epsBC_arc} on ${\cal L}_0$ at order $\varepsilon^1$,
\begin{equation}
    \partial_Y W_1(X,0) = C_1,\quad |X|<1.
\end{equation}
The solution to this boundary value problem was expressed by \citet{Philip_1972} as the imaginary part of a holomorphic function $f_P(Z)$
\begin{equation}\label{eq:solW1}
    W_1(X,Y)=-C_1 W_P(X,Y)=-C_1 \Im[f_P(X+iY)].
\end{equation}
With the Jacobi elliptic functions $\text{cn}(u,k)$, $\text{cd}(u,k)$ and the complete elliptic integrals of the first kind $K(k)=\int_0^{\pi/2}\frac{d\theta}{\sqrt{1-k^2\sin^2\theta}}$, $K'(k)=K(\sqrt{1-k^2})$, the desired function reads \citep[Eqns. (8.2), (8.7)]{Philip_1972}
\begin{equation}\label{eq:fPhilip}
    f_P(Z) = \frac{D}{K'(k_1)} \text{cn}^{-1}\left[\frac{\text{cn}(B^{-1}K(k)Z,k)}{\text{cn}(B^{-1}K(k),k)},k_1\right]-Z,
\end{equation}
where the elliptic moduli $k$ and $k_1$ are such that
\begin{align}
    &K'(k)/K(k)=D/B, \label{eq:def_k}\\
    &k_1=k\,\text{cd}(B^{-1}K(k),k). \label{eq:def_k1}
\end{align}
The contribution $C_1$ to the dimensionless stress along the gas-liquid interface is obtained from the mass conservation condition \eqref{eq:epsIntBC_arc}, for which we need the integral of $W_1$ on the $x$-axis. With 
\begin{equation}\label{eq:def_B1}
    B_1=D\,K(k_1)/K'(k_1)
\end{equation}
one obtains \citep[Eq. (3.16)]{Philip_1972b}
\begin{equation}\label{eq:Average_W10_Philip}
    \overline{W_P}(B,D) = \frac{1}{2B}\int_{-1}^1 \Im[f_P(X)] dX = D(1-B_1/B).
\end{equation}
Thus from \eqref{eq:epsIntBC_arc} with \eqref{eq:perturbationInterface} the first order contribution to the interfacial shear stress becomes
\begin{equation} \label{eq:solC1}
    C_1(B,D) = \frac{4}{3} \frac{1}{2B\overline{W_P}(B,D)} = \frac{2}{3} \frac{1}{DB(1-B_1/B)}.
\end{equation}
We note that in the limiting case where the flow is driven by constant shear stress far from the interface
\begin{equation}
    \lim_{D\to\infty} f_P(Z) = \frac{1}{\alpha} \arccos\left(\frac{\cos(\alpha Z)}{\cos\alpha}\right)-Z,\quad \alpha = \frac{\pi}{2B}
\end{equation}
and correspondingly for a single groove in the lower plane,
\begin{equation}
    \lim_{B,D\to\infty} f_P(Z) = \sqrt{Z^2-1}-Z,
\end{equation}
which agrees with the result obtained in \cite{Baier_2022}. The corresponding contributions to the shear stress on the gas-liquid interface are
\begin{equation}
  \lim_{D\to\infty} C_1(B,D) = \frac{4}{3\pi}\frac{\alpha^2}{\log(\sec\alpha)},\qquad
  \lim_{B,D\to\infty} C_1(B,D) = \frac{8}{3\pi}.
\end{equation}
Note that $C_1>0$ for all $B$ and $D$.

\subsection{Stress and effective slip: Lorentz reciprocal theorem}\label{sec:Stress_EffectiveSlip}

The average stress necessary for moving the upper wall at a given velocity with respect to the lower wall can be obtained by use of the Lorentz reciprocal theorem \citep{Kim_2005}, 
\begin{equation}\label{eq:LorentzReciprocal}
    \int_{\partial V} \mathbf{n}\cdot\boldsymbol{\tau}\cdot\hat{\mathbf{u}}\,dA
    = \int_{\partial V} \mathbf{n}\cdot\hat{\boldsymbol{\tau}}\cdot\mathbf{u}\,dA,
\end{equation}
where $\mathbf{n}$ is a unit normal pointing into the domain, relating stresses and velocities in an integral over the boundary $\partial V$ of the domain $V$. Here $\boldsymbol{\tau}=p\mathbf{1}-\mu(\boldsymbol{\nabla}\mathbf{u}+(\boldsymbol{\nabla}\mathbf{u})^T)$ is the stress tensor in an incompressible Newtonian fluid with velocity $\mathbf{u}$ and pressure $p$ obeying the Stokes equation $\boldsymbol{\nabla}\cdot\boldsymbol{\tau} = 0$ in some domain $V$ subject to certain conditions on its boundary $\partial V$. The reference flow of velocity $\hat{\mathbf{u}}$ and stress $\hat{\boldsymbol{\tau}}$ obeys the same conditions inside $V$ but solves for different conditions on the boundary.

We take as the main flow the velocity field $\mathbf{u} = a\dot{\gamma} W(X,Y)\mathbf{e}_z$ obtained by the domain perturbation method with expansion \eqref{eq:perturbationExpansion} and defined in the rectangular domain $\Omega_0$. As reference flow we take the Couette flow $\hat{\mathbf{u}} = \dot{\gamma}_\infty y \mathbf{e}_z$ between parallel plates separated by a distance $d$ with no-slip boundary conditions. We now apply \eqref{eq:LorentzReciprocal} in $\Omega_0$ in order to obtain the average stress on the upper wall due to the velocity field $W(X,Y)$. We note that there is no contribution to either side of \eqref{eq:LorentzReciprocal} from the symmetry boundaries $\Sigma_\pm$ as $\mathbf{n}\cdot\boldsymbol{\tau} = \mathbf{0} = \mathbf{n}\cdot\hat{\boldsymbol{\tau}}$ there. To the integral on the left hand side of \eqref{eq:LorentzReciprocal} only the top wall contributes with $-\int_{-b}^b \tau_{zy}(x,d) (\dot{\gamma}d) dx$ since $\hat{\mathbf{u}}$ vanishes on the lower wall. The right hand side integral has contributions from both the bottom and top wall with $\int_{-a}^a (-\mu\dot{\gamma}) w(x,0)dx - \int_{-b}^b (-\mu\dot{\gamma}) (\dot{\gamma}d) dx$. Introducing the dimensionless average velocity on the $x$-axis
\begin{equation}\label{eq:slipLengthIntegral}
    \Delta(B, D) = \frac{1}{2B}\int_{-B}^B W(X,0) dX
\end{equation}
and the dimensionless stress required to move the upper wall, 
\begin{equation}\label{eq:stressAverage}
    \overline{T}(B,D) = \frac{1}{(-\mu\dot{\gamma})}\left[ \frac{1}{2b}\int_{-b}^b \tau_{zy}(x,d)dx \right] = \frac{1}{2B}\int_{-B}^B \partial_y W(X,D) dX,
\end{equation}
we obtain from the Lorentz reciprocal theorem
\begin{equation}\label{eq:StressUpperPlate}
    \overline{T}(B,D) = 1-\frac{\Delta(B,D)}{D}.
\end{equation}
Finally, using \eqref{eq:epsIntBC_arc} and \eqref{eq:epsBC_arc}, we evaluate the average velocity on the $x$-axis in the perturbation expansion as
\begin{equation}
\begin{split}\label{eq:Delta}
    \Delta(B, D) &= \frac{1}{2B}\int_{-B}^B \left(\varepsilon W_1(X,0)+\varepsilon^2 W_2(X,0)\right) dX \\
    &= \frac{1}{2B}\int_{-B}^B \left(\varepsilon \left[-H_1(X)\right]+\varepsilon^2 \left[-H_1(X)\partial_Y W_1(X,0)\right]\right) dX\\
    & = -\frac{1}{2B}\int_{-B}^B \left(\varepsilon H_1(X)+\varepsilon^2 C_1 H_1(X)\right) dX 
    = -\frac{2}{3B}\left(\varepsilon + \varepsilon^2 C_1 \right)
\end{split}
\end{equation}
Note that since \eqref{eq:epsIntBC_arc} constrains the average of $W_2$(X,0), we have obtained $\Delta$ at order $\varepsilon^2$ without needing to obtain the velocity field at this order.

\subsubsection{Effective slip length}
We define the non-dimensional effective slip length $\Lambda$ in the system with structured plates separated at a distance $D$ by setting the stress \eqref{eq:StressUpperPlate} necessary to move the upper plate equal to the stress in a reference system of parallel flat no-slip plates separated at a distance $D+\Lambda$. Thus $\overline{T}(B,D) = \frac{W(X,D)}{D+\Lambda} = \frac{D}{D+\Lambda}$ from which we obtain as definition for the effective slip length
\begin{equation}\label{eq:defSlipLength}
    \Lambda(B,D) = \frac{D}{\overline{T}(B,D)}-D.
\end{equation}
For the analytical calculation, using \eqref{eq:StressUpperPlate} and \eqref{eq:Delta}, this becomes
\begin{equation}\label{eq:analyticSlipLength}
    \Lambda(B,D) = \frac{\Delta(B,D)}{1-\Delta(B,D)/D} = -\left(\frac{1}{D} + \frac{3B}{2\varepsilon(1+\varepsilon C_1(B,D))}\right)^{-1},
\end{equation}
which for large plate separation $D$ simplifies to $\Lambda \simeq \Delta$. Sufficiently far from the surface, at distances $y \gg 2b$, the detailed influence of the structure at the lower wall becomes negligible and the flow appears as a simple shear flow with shear rate $\dot{\gamma}$ and velocity $w\sim\dot{\gamma}(y+a\Lambda)$, indicating that this definition of the effective slip length is in agreement with the definition in case of unbounded shear flow over a structured plate \citep{Rothstein_2010}. Note that from \eqref{eq:Delta}, $\Delta=-\frac{2}{3B}\left(\varepsilon + \varepsilon^2 C_1\right)$, and hence $\Lambda$ is negative for positive deflection $\varepsilon$, corresponding to an increase in drag compared to a flow over an unstructured no-slip plate. We will come back to this in the discussion below.

\subsection{Numerical calculations}\label{sec:finiteElement}

In order to go beyond the limitations of the domain perturbation method, the Laplace equation, \eqref{eq:noDim_Laplace}, is also solved numerically with the commercial finite-element solver COMSOL Multiphysics (version 6.1, COMSOL AB, Stockholm, Sweden), using the 'Coefficient form PDE' interface. Due to the mirror symmetry with respect to reflection at the $y$-axis, these calculations are performed in a rectangular domain $0<X<B$, $0<Y<D$, with a circular-arc section, corresponding to the deflected gas-liquid interface, added or removed above or below the $X$-axis at $X<1$, coinciding with the region to the right of the $y$-axis on figure \ref{fig:sketch}(b). As was shown in \citep{Baier_2022}, it suffices to prescribe the integral conservation equation \eqref{eq:dimIntegralMassConservationSurfaceFluid} as a constraint on the circular arc for modelling the incompressible surface fluid, and a Dirichlet condition, $W=0$, enforces the no-slip condition on the rest of the bottom surface. A constant velocity $W = D$ is applied on the top surface at $Y=D$, and a vanishing shear rate, $\partial_X W = 0$, is assumed on the left and right edges at $X=0$ and $D$, corresponding to a symmetry condition. We discretize the domain using quadratic Lagrange elements on a triangular mesh with cells of size $h_B=0.025$ away from the surface and $h_S=h_B/5$ on the circular arc ${\cal S}$ and the solid section ${\cal L}_+$ of the bottom wall, with a maximal element growth rate of 1.01. Using Richardson extrapolation \citep{Pomeranz_2011} to assess the grid dependence, it was verified that in the parameter range under investigation the velocity at the center of the interface $W(0,\varepsilon)$, the effective slip length $\Lambda$ and the interface shear rate $C$ differ by at most 1\,\% from the extrapolated results.

Additional calculations were performed in which the boundary condition on the circular arc ${\cal S}$ was replaced by a no slip condition, $W|_{\cal S} = 0$, corresponding to solid protrusions, or a shear-free condition, $\mathbf{n}\cdot\boldsymbol{\nabla}W|_{\cal S} = 0$, corresponding to uncontaminated gas-liquid interfaces.

\section{Results and discussion}

\begin{figure}
	\centering
	\includegraphics[width=0.9\textwidth]{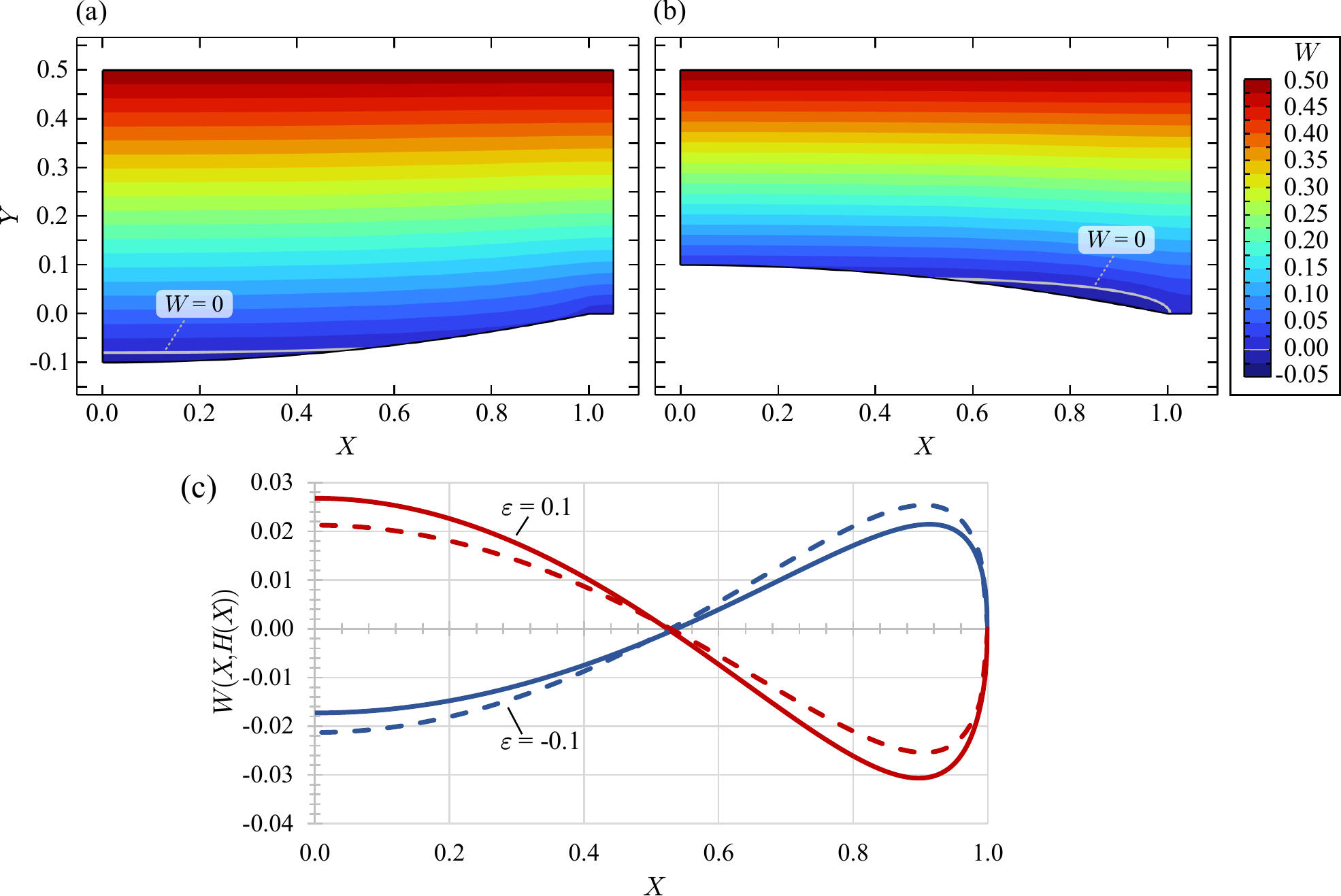}
	\caption{\label{fig:velocity} Velocity $W(X,Y)$ for (a) $\varepsilon=-0.1$, (b) $\varepsilon=0.1$ at plate separation $D=0.5$ and domain width $B=1.05$. The grey line indicates the iso-contour $W(X,Y)=0$. (c) Velocity on the interface, $W(X, H(X))$ (solid line), and analytical approximation, $\widetilde{W}(X)$ (\mbox{eq.~\eqref{eq:Wtilde}}, dashed line) for interface deflections $\varepsilon=\pm 0.1$.}
\end{figure}

In figure \ref{fig:velocity} we show exemplary plots of the velocity $W(X,Y)$ for a small plate separation $D=0.5$ and narrow pillars between grooves, $B=1.05$, obtained by numerical calculation. Qualitatively, the same picture presents itself for other values of the plate separation $D$ and domain widths $B$. In figure \ref{fig:velocity}(a) the interface is curved downward by a maximal deflection $\varepsilon=-0.1$, while in \ref{fig:velocity}(b) the deflection is upward with $\varepsilon=0.1$. A striking feature in these graphs are the small values of the velocity at the surfactant-laden interface, with a nearly linear rise of velocity towards the moving upper wall. As mentioned, since the grooves have a finite length, surfactant that is transported along the groove in the same direction as the movement of the upper plate must be transported in opposite direction on other parts of the interface, as encapulated in the integral mass conservation \eqref{eq:dimIntegralMassConservationSurfaceFluid}. To highlight these regions of recirculating flow we show the isoline $W(X,Y)=0$ in gray with flow in opposite direction of the movement of the upper plate below this line. This is more clearly seen in figure \ref{fig:velocity}(c) where the velocity $W(X,H(X))$ on the interface from the numerical calculation is shown as solid lines for $\varepsilon=\pm 0.1$. An interface deflected towards the upper wall results in co-flow at the center of the interface with the corresponding backflow towards the edges of the groove. The opposite picture presents itself for a negative deflection of the interface. 

In order to put the observed velocities into perspective, it should be noted that flow over a flat shear-free interface can be expressed by the imaginary part of \eqref{eq:fPhilip}, giving velocities at the center of the groove of order 1 for sufficiently large $D$ (and of order $D$ for very small plate separation). Flow over an interface covered by an incompressible surfactant is thus much slower and therefore more akin to flow over a solid surface with protrusions in or out of the plane of the surface. In particular, in the case of a flat interface we already noted that the interface remains completely immobilised when an incompressible surface fluid is present.

In order to compare with the analytic expression \eqref{eq:perturbationExpansion} for the velocity field on the interface, we expand $W(X,\varepsilon H_1(X))$ to first order in $\varepsilon$ as
\begin{equation}\label{eq:Wtilde}
    \widetilde{W}(X) = \varepsilon H_1(x) + \varepsilon W_1(X,0)
\end{equation}
with $H_1(X)$ from \eqref{eq:perturbationInterface} and $W_1(X,0)$ from \eqref{eq:solW1}. Not only is this expression exact to the same order in $\varepsilon$ as $W(X,\varepsilon H_1(X))$ itself, but it is also more in line with the approximation \eqref{eq:epsIntBC_arc} used as the projected integral boundary condition on the interface. The velocities $\widetilde{W}(X)$ are shown as dashed lines in figure \ref{fig:velocity}(c) for the same configurations as in the numerical calculations. As can be seen, while the order of magnitude for the velocity is captured by the analytical expression, there is a noticeable difference between the numerical and analytical values even for this moderate value of the deflection. This is also apparent in the fact that at this order in the expansion the analytical approximation $\widetilde{W}(X)$ is symmetric in $\varepsilon$, while the numerical results show that this symmetry is only approximate. This is to be expected, as we have only obtained the velocity as a linear approximation around $\varepsilon = 0$. However, since our main interest is the average shear stress $\overline{T}$, or equivalently the apparent slip length $\Lambda$, which is known to second order in $\varepsilon$, better agreement is to be expected for these quantities even at moderate deflections. Nevertheless, the analytic expression already predicts a monotonic increase in the velocities at the interface with increasing deflection and the small values compared to a shear-free interface.

\begin{figure}
	\centering
	\includegraphics[width=\textwidth]{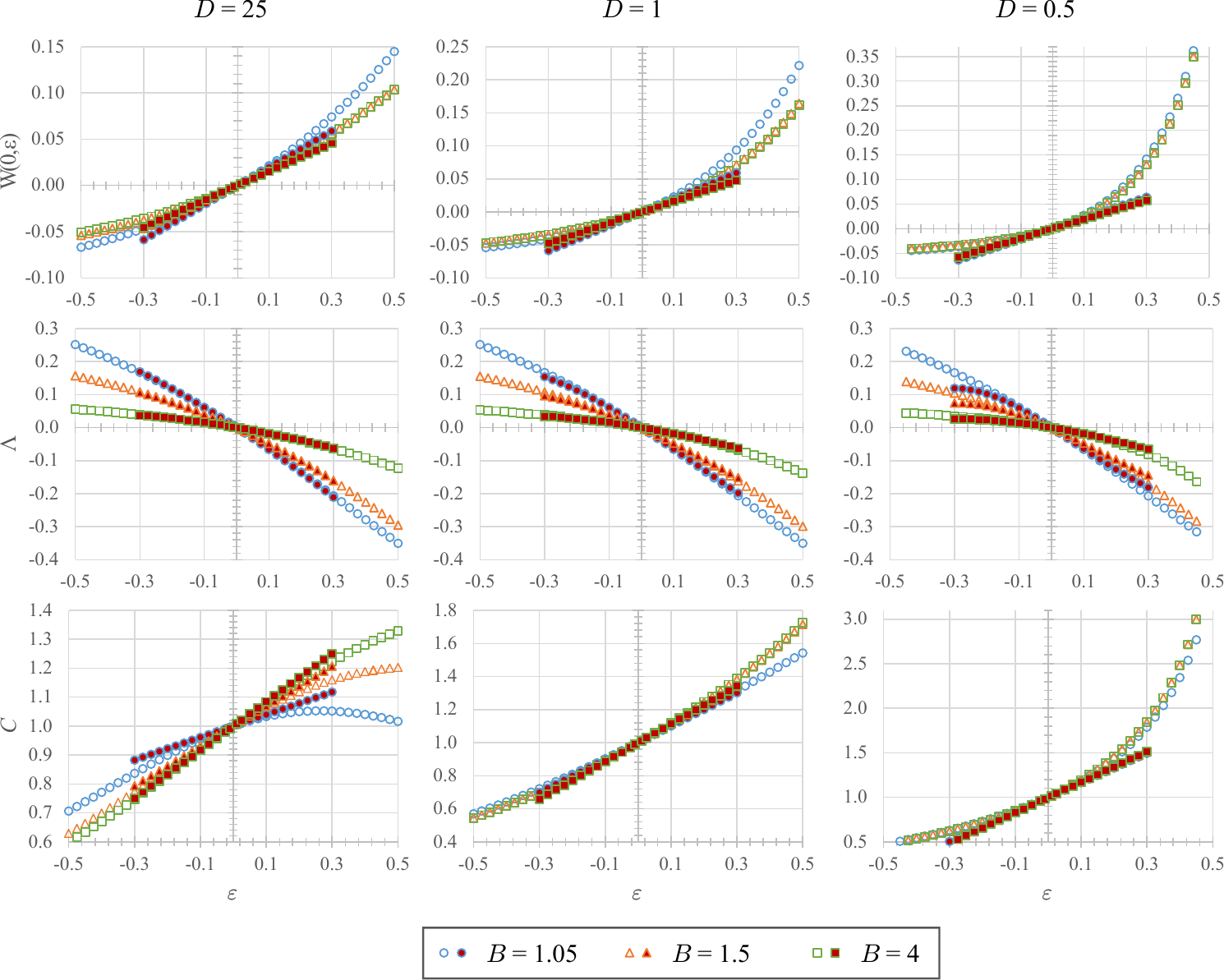}
	\caption{\label{fig:WLambdaC} Velocity at the center of the interface, $W(0,\varepsilon)$, effective slip length $\Lambda$ and interfacial shear rate $C$ for plate separations $D=25$, 1 and 0.1. Open symbols correspond to numerical calculations with the values of $B$ listed in the legend, while filled symbols correspond to the analytical approximations \eqref{eq:Wtilde} with \eqref{eq:solW1}, \eqref{eq:solC1} for $W$, \eqref{eq:analyticSlipLength} with \eqref{eq:solC1} for $\Lambda$ and \eqref{eq:solC1} for $C$.} 
\end{figure}

The dependence of the interfacial velocity scale, encoded in $W(0,\varepsilon)$ at the center of the meniscus, on $\varepsilon$ is shown in the first row of figure \ref{fig:WLambdaC} for $D=25$, 1 and 0.5, respectively, and for variable widths $B=1.05$, 1.5 and 4 of the unit cell. As extreme values for the interface deflection are unlikely to become experimentally accessible, we restrict the analysis to deflections between $\varepsilon=\pm 0.5$ for $D=25$ and 1, and $\varepsilon=\pm 0.45$ for $D=0.5$. For comparison, the analytical expression $\widetilde{W}(0)$ is plotted with filled symbols between $\varepsilon=\pm 0.3$, while the numerical values are shown using open symbols. It is apparent that there is some nonlinearity in the numerically obtained magnitude of the interfacial velocity with the deflection, particularly pronounced at small plate separations $D$, explaining the relatively poor performance of the analytical expression \eqref{eq:Wtilde} even at moderate deflections under these conditions. However, qualitatively the most striking feature, namely the small magnitude of the interfacial velocity scale, is still captured. Note that for positive deflections the velocity scale increases with decreasing plate separation, but still remains much smaller than the velocity of the upper plate even for the extremely small gap at $D=0.5$ and $\varepsilon=0.45$. Also note that there is little difference between results obtained for wide spacing between the grooves, $B=1.5$ and $B=4$, where the velocity curves lie practically on top of each other, indicating little influence between neighboring grooves on the flow. Naturally, for small plate separation $D$ this cross-talk is even reduced.

The main property of interest in this investigation is the average stress at the upper plate, as this is more directly accessible to experiments in a viscosimeter than the interfacial velocity. More specifically, we are interested in the deviation between the necessary applied stress in this configuration compared to the corresponding set-up with flow between unstructured plates at the same separation $D$, and will therefore focus on the effective slip length $\Lambda$ defined by \eqref{eq:defSlipLength}. For the same geometric configurations as for the interfacial velocities, the effective slip length is shown in the second row of figure \ref{fig:WLambdaC}. Numerically, $\Lambda$ is derived directly from the average shear at the upper surface together with the definition  \eqref{eq:defSlipLength}, and the corresponding values are shown as open symbols. The analytical approximation for $\Lambda$ is obtained from \eqref{eq:analyticSlipLength} using \eqref{eq:solC1} and shown as filled symbols. As can be seen, there is excellent agreement between the numerical and analytical values even up to $|\varepsilon|=0.3$ for sufficiently large plate separation, with the analytical expression nicely capturing the nonlinear behaviour as function of $\varepsilon$. Nevertheless, as for the velocities, the agreement is significantly poorer for small plate separation $D=0.5$. It is interesting to note that the numerically obtained effective slip lengths $\Lambda$ are nearly independent of $D$ within the range of parameters investigated. However, the main take-away is that the slip lengths are negative for positive interface deflection and positive for negative deflection. Thus, for $\varepsilon>0$ the stress at the upper wall is larger than required for establishing the flow between unstructured plates at the same separation $D$. Due to the small velocities at the interface this is to be expected and the conditions are rather similar to flow over a wall with solid protrusions in and out of channel. The situation is thus very different from the expectation one may have from flow over a striped superhydrophobic surface with clean interfaces. Obviously, a sufficient coverage of the interface with surfactants is strongly adverse to the expected slip-enhancement of superhydrophobic surfaces when bounded cavities are used, enforcing recirculation of surfactant on the interface.

Finally, we report the interfacial stress $C$ in the third row of figure \ref{fig:WLambdaC}. As the interfacial stress derives from the gradient in surface pressure, too large values may lead to breakdown of the interface layer \citep{Lee_2008} and may in some situations limit the applicability of the simple model for the incompressible surfactant layer used in the analysis. Numerically, $C=\partial_Y W(0,\varepsilon)$ is most easily evaluated at center of the interface, while the analytic results are derived from \eqref{eq:solC1}. Since in the present analysis $C$ is only evaluated to first order in the interface deflection $\varepsilon$, the corresponding expression is limited to small deflection, but nevertheless captures the order of magnitude of the interfacial stresses. Nevertheless,  significantly larger stresses can occur for small gaps and large positive deflection. An interesting feature is the non-monotonicity of $C$ as function of $\varepsilon$ for large gaps $D=25$ and $B=1.05$ due to the interaction between velocity fields at neighboring cavities.

\begin{figure}
	\centering
	\includegraphics[width=0.9\textwidth]{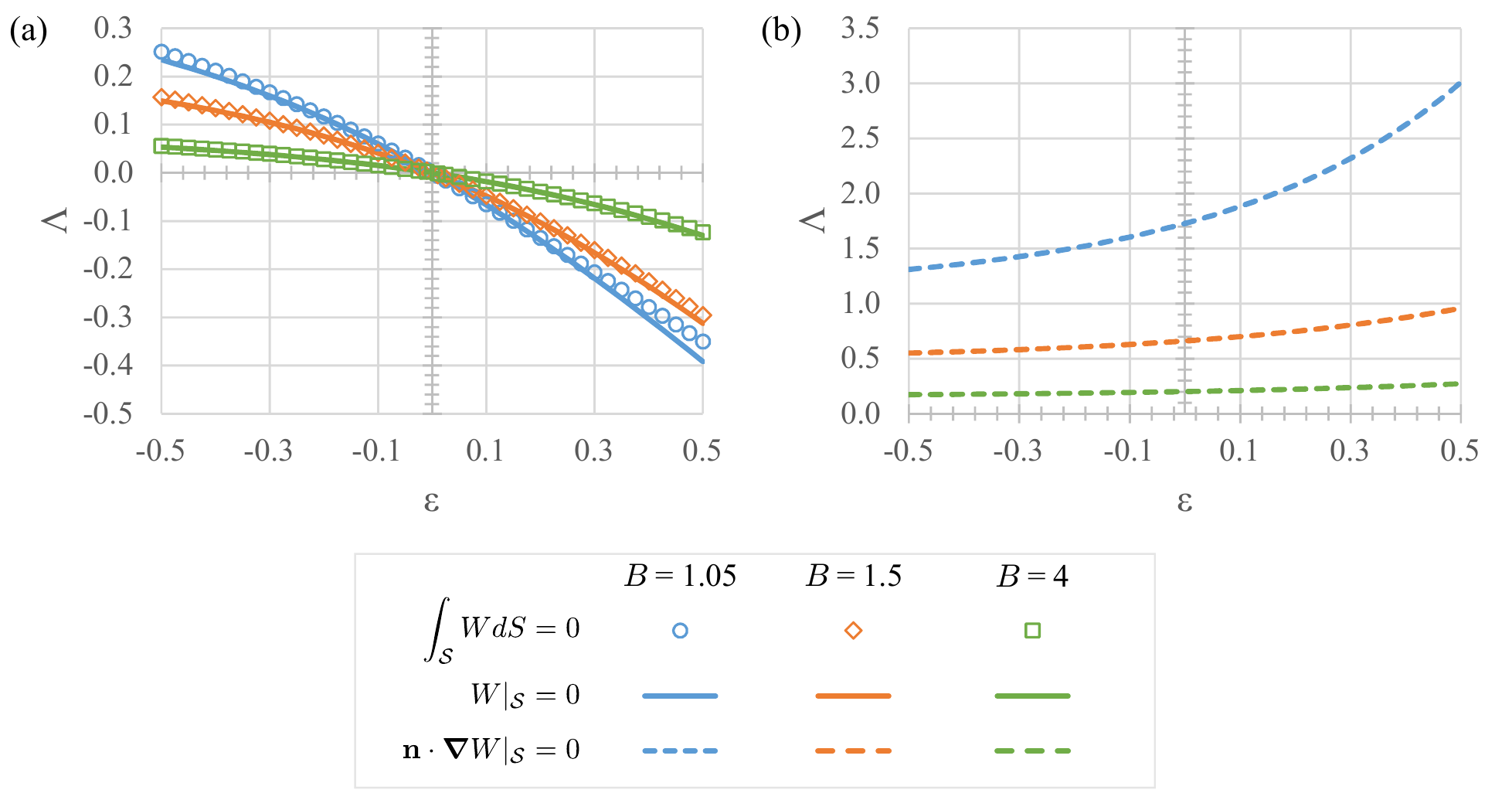}
	\caption{\label{fig:noSlipFreeSlip} Effective slip length $\Lambda$, derived from numerical calculations, for plate separation $D=25$ and the values of $B$ listed in the legend corresponding to (a) interfaces $\cal S$ covered with incompressible surfactants (open symbols), solid protrusions (solid lines) and (b) shear-free interfaces (dashed lines).} 
\end{figure}

As we have seen, in the case where the gas-liquid interfaces are covered by an incompressible surfactant phase, the velocities at the interface remain far below the values expected at an uncontaminated gas-liquid interface with a shear-free boundary condition. It is thus instructive to compare the results for the effective slip length obtained for the surfactant covered interface with results where the interface remains stationary, corresponding to solid no-slip protrusions, or where the interface is shear-free, corresponding to uncontaminated gas-liquid interfaces. We focus here on a large plate separation $D=25$, but analogous results are obtained in the cases with $D=1$ and 0.5 considered before. Figure \ref{fig:noSlipFreeSlip}(a) shows a direct comparison between the effective slip length for flow over an array of interfaces covered by an incompressible surfactant phase and flow over corresponding solid protrusions. As expected from the small interface velocities observed at the surfactant covered interface, there is little difference between the slip length obtained in these cases. Nevertheless, the values for the slip length in the case of surfactant covered interfaces consistently lie slightly above the corresponding values for the stationary interface, indicative of the mobile nature of the interface in the former case. Note that these findings are in agreement with the observations recently made by \citet{Rodriguez_2023} for flow over such surfaces in the dilute (large $B$) limit of large separation between cavities. For reference, the effective slip length in the case of an uncontaminated interface, approximated by a shear-free surface, is shown in figure \ref{fig:noSlipFreeSlip}(b). As expected, in particular for small $B$, the reduction in drag at the partially stress-free structured plate leads to a consistently positive and significantly larger slip length than in the other two cases. Nevertheless, we remark that in other cases such as flow over surfaces with an array of bubbles or transverse gas-filled ridges, the effective slip length can even become negative for large enough protrusions of the bubbles into the channel despite a shear-free gas-liquid interface \cite{Steinberger_2007, Hyvaluoma_2008, Davis_2009, Ng_2011, Karatay_2013, Haase_2013, Lee_2016}.

\section{Conclusion and Outlook}

We have investigated the influence of an incompressible surfactant phase at the gas-liquid interface on the apparent slip in shear driven flow over a superhydrophobic surface containing a regular array of gas-filled grooves. Since the gas-liquid interfaces are bounded by the edges of the cavities and the surfactant is assumed to be insoluble in the liquid, mass conservation within the surfactant phase demands that the net flow of surfactant along the grooves vanishes. For flat gas-liquid interfaces this leads to complete immobilisation of the interface, while in the case of curved interfaces a recirculating flow pattern appears. For positive deflection of the interface into the fluid region the flow is in the direction of the applied shear stress at the centerline of the groove with recirculating flow at its edges, and vice versa for negative deflection below the plane of the structured wall. In all cases the maximal velocity at the interface remains far below the velocity one expects for a clean, approximately shear-free interface within the range of moderate interface deflection studied. Indeed, compared to the case of a planar no-slip surface, a higher (or lower) shear rate is necessary to drive the flow along the superhydrophobic surface with interfaces curving into (or out of) the fluid domain due to the presence of the incompressible surfactant phase, and this is reflected in a negative slip length. In this respect flow over the surfactant laden interfaces thus rather resembles a situation of flow over a surface with corresponding no-slip protrusions extending above or below its plane. For large enough plate separations $D$ the derived analytic expression for the effective slip length poses an excellent approximation for the numerically obtained values even at moderate deflections $\varepsilon$.

Note that in an even stricter sense the argument for immobilization of the interface also extends to flow in transverse direction over the array of superhydrophobic surfaces covered by a sufficient amount of insoluble surfactant. For a flat interface this was investigated in \citep{Baier_2021} and \citep{Mayer_2022}. In this case, an explicit balance between the applied viscous shear stress and the Marangoni stress within the surfactant film was performed to find the surfactant concentrations at the interface. At sufficiently large Marangoni number, or correspondingly large surface coverage, and sufficiently large P{\'e}clet number, the interface becomes effectively immobilized. The same also applies to transverse flow over curved interfaces, where again the surfactant distribution adjusts such that Marangoni stress balances the shear stress for flow over a corresponding immobilized surface. This requirement is weaker than, but includes the limiting case of an effectively incompressible surfactant phase. Thus, slow flow at any angle over a superhydrophobic surface containing an array of gas-filled grooves with an interface covered by an incompressible surfactant phase, can be decomposed into its tangential component investigated here and a corresponding transverse component of flow over corresponding solid protrusions, and will have a tensorial character similar to anisotropic Poiseuille flow between textured plates \citep{Stroock_2002, Bazant_2008, Kamrin_2010}.

An important criterion for the design of surfaces aimed at near-wall drag reduction by incorporating gas-filled cavities can be derived from the present analysis, as sufactants are ubiquitous in such applications. Since surfactants can stack up at the edges of bounded cavities it becomes advantageous to design surfaces containing effectively unbounded interfaces, such as superhydrophobic arrays of posts or pillars in the partially wetted Cassie state. However, an unbounded gas film may become more easily drained from the surface, leading to a collapse into the fully wetted Wenzel state. An alternative approach may therefore be a surface design containing regions where surfactant can stack up without affecting the flow and from where surfactants can be diverted, removing them from the functional sections of the interface.

\begin{acknowledgments}
	I am indebted to Steffen Hardt for his valuable input during many stimulating discussions.
\end{acknowledgments}




\bibliography{references}

\end{document}